\documentstyle[preprint,eqsecnum,aps]{revtex}

\begin{document}
\draft

\title{Possible disordered ground states for close-packed polytypes
and their diffraction patterns}

\author{Jaichul Yi and Geoff S. Canright}

\address{
Department of Physics and Astronomy, University of Tennessee\\
Knoxville, TN 37996-1200}

\maketitle
\begin{abstract}

It has recently been shown that one-dimensional Ising problems can have
degenerate, disordered ground states (GSs) over a finite range of coupling
onstants, ie, without `fine tuning'. The
disorder is however of a special kind, consisting of arbitrary mixtures
of a short-period structure and its symmetry-degenerate partner or
partners.
In this exploratory study,
we assume that the energetics of close-packed polytypes
can be represented by an Ising Hamiltonian (which includes, in principle,
all terms allowed by symmetry),
and that (for
simplicity) the close-packed triangular layers are unfaulted and
monatomic. We then calculate the diffraction patterns along the stacking
direction for the various possible kinds of disordered GSs.
We find that some disordered GSs give
diffraction patterns which are only weakly distinguished from their
periodic counterparts, while in others the disorder is more clearly evident
beneath the delta functions. Finally, in some cases, the
long-ranged order of the layers is destroyed
by the interference among the
short-period structures and their symmetry-related partners,
giving a diffraction spectrum which is purely continuous.

\end{abstract}


\narrowtext
\section{Introduction}
For one dimensional $k-$state models it is known that the
ground state is `almost always' periodic.
Radin and Schulman{\cite{rs}} showed that, for such models, any nondegenerate
ground state is periodic, and that,
for the degenerate case, there always exists at least one periodic
ground state.
In each case the period is at most $k^r$
where $r$ is the range of interaction.
For $k=2$ (Ising model) Teubner{\cite{teubner}} obtained the same results
using the directed graph $G_r^{(k)}$
(called the de Bruijn diagram after N. G. de Bruijn's paper\cite{bruijn}).
Recently Canright and Watson (CW){\cite{cw}}
considered the mathematically `exceptional' (but physically
unexceptional) case of Hamiltonians
constrained by symmetry. CW showed that, for many values of $k$ and
$r$, the restriction to symmetric Hamiltonians leads to the possibility
of degenerate and
disordered GSs over a finite fraction of coupling-parameter space.
(This finite fraction is of course negligible in the higher-dimensional
space of all possible Hamiltonians, unconstrained by symmetry.)
That is, if one allows the physically unexceptional `fine tuning' of
parameters arising from symmetry, then in most cases one can find
degenerate and disordered GSs without any further fine tuning
of parameters. The disorder arises since, in such cases, there are
degenerate periodic states (phases) such that the energy of a domain
wall between the degenerate phases is zero. Hence any
arbitrary (and so in general aperiodic) mixture of the degenerate states
is also a ground state.

Such ground states have a finite entropy per spin, and so suggest the
possibility of (weak) exceptions to the third law of thermodynamics
\cite{3law}.
It is therefore
of interest to inquire whether CW's theoretical results may be
applied to any real physical systems. Here we consider close-packed
polytypes\cite{verma_krishna}, which are well modeled as classical Ising
($k=2$) chains. As a first step in the study of this special kind of
one-dimensional disorder, we examine in this paper
the possible kinds of diffraction patterns (along the stacking axis)
for the disordered GSs found by CW. Our goal is to try to see
how the constrained disorder described above (ie, mixtures of two
distinct stacking sequences which are related by symmetry)
may be realized in the experimentally accessible form of the diffraction
pattern.

\section{Directed Graphs}

Our approach, like that of CW, relies on the representation of a
Hamiltonian $H_r^{(k)}$ (with $r$ the range of interaction, and $k$ the number
of states per site) as a directed graph $G_r^{(k)}$. Hence,
in this section, we provide a brief description
of this representation, including the modifications introduced by CW to
represent the effects of symmetry.

$\mbox{From}$ here on we will
restrict our attention on the Ising model ($k=2$), and so drop the
superscript $k$ everywhere.
An infinite Ising chain with
interaction range $r$ can be viewed as a successive sequence
of spin configurations, each of length $r$.
There are $2^r$ such configurations; these become
the nodes ($\cal N$) of the graph $G_r$.
To complete the graph, two nodes ${\cal N}_1=\sigma_1...\sigma_r$
and ${\cal N}_2=\sigma_1^\prime ...\sigma_r^\prime$ are connected by an
arrow (directed arc) only if the last $(r-1)$
spins of ${\cal N}_1$ are identical
to the first $(r-1)$ spins of ${\cal N}_2$. This arc then represents a
transition ${\cal N}_1 \to {\cal N}_2$, effected by the addition of the spin
$\sigma_r^\prime$ to the chain (which we imagine as growing from the
left).
The $2^r$ nodes in $G_r$ are thus connected by $2^{r+1}$ arcs,
each of which can be labeled with $(r+1)$ sequential spin values.
The graph $G_r$ then represents the Hamiltonian $H_r$ as follows.
A unique
weight (energy cost resulting from adding a spin $\sigma_r^\prime$
to the chain) can be associated with each arc.
Any infinite Ising chain of spins
can be then represented as a path through the graph $G_r$, with the
energy of the chain being simply the sum of the energies (weights) of
the arcs in the path. Since the graph has a finite number of nodes,
any infinite path must visit at least one node more than once; hence,
ignoring boundary effects, such a chain must be a closed cycle in $G_r$.
Furthermore, if we define a simple cycle (SC) as a non-decomposable
(ie, non-self-intersecting) cycle, then all the cycles in $G_r$ can be
uniquely decomposed into SCs.

The general periodicity of the ground state can now be understood
in terms of SCs. The gound state is the repetition of that SC which has
the minimum energy per spin, and the period of any SC is $\le$ $2^r$
(the number of nodes of $G_r$).
In the case that the parameters in $H_r$ are `fine tuned'
to precise values (as can occur from symmetry),
there can be more than one SC with
the least energy per spin in the graph $G_r$---that is, there can be
degenerate ground states which are related by symmetry.
We are interested here in the two symmetries $S$ (spin inversion) and
$I$ (space inversion). These symmetries force the degeneracy of {\it pairs\/}
of SCs in $G_r$. Now assume that such a pair has the lowest energy per spin.
If these two minimal-energy SCs share a node,
then the GS is infinitely degenerate, since it includes arbitrary
mixtures of the two SCs.
If on the other hand they do not share any nodes, it is evident
that jumping from one cycle
to another costs energy, and so gives a configuration which is not a GS.
CW found that the former case (infinitely degenerate GS, arising from
a pair of minimal-energy SCs which share a node or nodes)
occurs for many values
of $k$ and $r$, assuming only $S$ or $I$ symmetry. We will follow their
terminology and call such a pair a `D-pair'. That is, a D-pair is a pair
of SCs in $G_r$ which (i) can be minimal-energy configurations for a
range of parameter values in $H_r$; (ii) are related by symmetry and hence
degenerate; (iii) share one or more nodes.

For the Ising model (more precisely for even $k$) CW found that
$S$ symmetry alone never gives rise to disordered ground states
(D-pairs).
For the case of $I$ symmetry,
CW showed that (again for $k=2$) D-pairs do exist,
but only for $r\ge 5$.
Combining the two symmetries (denoted $S+I$ symmetry), CW found that
the Ising case has D-pairs only for $r\ge 6$.

These results may seem somewhat surprising, from the following point of
view. It is easy to find SCs of the graph $G_r$ which are related by
symmetry and hence satisfy (ii) above, while also satisfying (iii).
Hence one might think that D-pairs should be ubiquitous.
However, it turns out that the imposition of symmetry often makes
satisfaction of (i) impossible, even as it enforces (ii), for a pair of
SCs. Hence CW turned to modified graphs $\hbox{}^XG_r^{(k)}$
(where $X$ is the symmetry $S$, $I$, or $S+I$) whose SCs always satisfy
(i). For the polytype problem we will concentrate on $\hbox{}^{S+I} G_r$.

The graph $\hbox{}^{S+I} G_r$ is most readily constructed (for details see CW)
by first operating on
$G_r$ with $S$ (giving $\hbox{}^S G_r$), and then operating on the latter with
$I$. Similarly, a SC of $\hbox{}^{S+I} G_r$ is mapped to its counterparts in
$G_r$ by reversing this sequence: first undoing $I$ (hence `unfolding' the SC
into $\hbox{}^S G_r$), then undoing $S$.
This unfolding of a SC of $\hbox{}^{S+I} G_r$ will
yield one, two, or more generally four SCs of $G_r$.
We are interested in those SCs of $\hbox{}^{S+I} G_r$ which,
upon unfolding, yield multiple node-sharing cycles.
Although the precise identification of
distinct pairs may be complicated by the simultaneous
presence of two symmetries, such unfolded
cycles are the analogs of the D-pairs identified by CW; we will use
the same term for these (multiple, unfolded) cycles in $G_r$.

It is useful to classify the D-pairs into topological types.
CW found two topological SCs of any graph of the form $\hbox{}^IG$
(which includes $\hbox{}^{S+I} G_r$) which represent D-pairs in $G$.
Expansion of one type gives any of four
topologically different types (I-IV) of D-pairs in $G_r$ (Fig.\ 1).
Expansion of the other type (which is found in $\hbox{}^{S+I} G_r$
only for $r\ge 8$) yields either a type IV or type V D-pair in $G_r$.

Let us briefly describe the five types of D-pairs.
First we may have a type I D-pair in which a cycle $cyc$ shares the
node $\cal N$ with its $I$ symmetry partner
$\mathop{cyc}\limits^{\longleftarrow}$ [Fig.\ 1(a)].
A type II D-pair is composed of $cyc$ and
its $SI$ symmetry partner
$\mathop{\overline{cyc}}\limits^{\longleftarrow}$ sharing
a node $\cal N$ [Fig.\ 1(b)].
Both type I and II D-pairs are always accompanied by another D-pair,
related to the first by $S$ symmetry (hence sharing the
node $\overline{\cal N}$).
Taking the length of the cycle $cyc$ (in the Ising or `01'
representation---see below) to be $T_{01}$, and assuming a random mixture of
the two node-sharing cycles, we get the entropy \cite{entropy}
of types I and II as $\ln 2$ per $T_{01}$ layers.

Fig.\ 1(c) and 1(d) shows type III and IV D-pairs.
We can think of these types as a pair of cycles,
which however share two nodes:
$\cal N$ and $\mathop{\overline{\cal N}}\limits^{\longleftarrow}$ for
type III, and $\cal N$ and $\mathop{\cal N}\limits^{\longleftarrow}$
for type IV.
It is clear from the figures that four distinct (but degenerate)
cycles may be formed from a type III or IV D-pair.
Hence the entropy for these is $\ln 4$ per $T_{01}$ layers.

In general, as with types I and II, types III and IV D-pairs
imply the existence of other D-pairs, related by $S$.
However such general (asymmetric) D-pairs only occur
for $r\ge 8$---a range which we have not
studied systematically. Hence we have instead shown the
symmetric cases, for which $S$(D-pair) = D-pair.
Such a symmetric pair will give cycles with period $T_{01}$
which is twice that of the `folded' D-pair in
$\hbox{}^{S+I} G_r$ (hence even).
These symmetric type III and IV D-pairs map to themselves
under either $S$ or $I$; hence [unlike Fig.\ 1(a) and 1(b)]
there are not other pairs implicit in Fig.\ 1(c) and 1(d).

In a type V [Fig.\ 1(e)] D-pair, all four symmetry related nodes
$\cal N$, $\mathop{\cal N}\limits^{\longleftarrow}$,
$\overline{\cal N}$, $\mathop{\overline{\cal N}}\limits^{\longleftarrow}$
are shared as shown.
Again, there are no other implicit pairs.
Since there are two choices of path at each shared node,
there are $2^4 = 16$ cycles (again, all degenerate)
represented by a type V D-pair, giving $S = \ln 16 = 4\ln 2$
per $T_{01}$ layers.

We note here that, besides the entropy, we can quantify the complexity
of the kinds of GS under study. Here we will use the definition of
Crutchfield and Young \cite{crutch}, which is trivially computed for our
D-pairs, since it relies upon the representation of chains as
probabilistic, finite-state automata. Thus we take the complexity as
$C = -\sum_n p_n \ln(p_n)$, where $n$ runs over the nodes of the
automaton (D-pair), and $p_n$ is the node probability. We can easily
obtain a general expression for $C$ for all the types of D-pair in
Fig.\ 1, as follows. Let $n_s$ be the number of shared nodes in the
D-pair (1 for types I and II, 2 for types III and IV, and 4 for type V),
and let $n_u = 2(T_{01}-n_s)$ be the number of unshared nodes.
Then $p_s = 1/T_{01}$ and $p_u = 1/(2T_{01})$. Thus the complexity of a
D-pair is $C_D = \ln(2T_{01}) - {n_s\over T_{01}} \ln 2$, which exceeds
the complexity $C_{per} = \ln(T_{01})$ of a periodic chain of the same
$T_{01}$ by $C_D - C_{per} = \ln 2(1-{n_s\over T_{01}})$.
Hence we see that a D-pair is less complex, by this definition,
than a periodic chain of period 2$T_{01}$; but its entropy is higher.

We wish to compute and study the diffraction patterns for these five
types of D-pairs. To do this we will translate Ising spin configurations
into an $ABC$-sequence of close-packed layers, using the standard
mapping between the two, as follows.
Any pair in the sequence $A\to B\to C\to A$ is
denoted by $1$ (or $+$, in H\"agg's notation\cite{hagg});
and a pair from the sequence $A\to C\to B\to A$ is a
$0$ ($-$).

We next introduce some notation, using an example for clarity.
One D-pair from $\hbox{}^{S+I} G_{6}$, for example, consists of $cyc=(0010111)$
(with Zhdanov symbol\cite{zhdanov} $cyc=(2113)_3$)
and its $SI$-symmetry partner
$\mathop{\overline{cyc}}\limits^{\longleftarrow}=(0001011)$.
(The shared node is $001011$.)
If the number of $1$'s and $0$'s in a cycle is denoted by $n_1$
and $n_0$, a parameter $\Delta$ can be defined by
$\Delta \equiv (n_1 - n_0) \pmod 3$.
Hence a cycle with $\Delta = 0$, if repeated periodically,
gives a hexagonal polytype, while one with $\Delta=\pm1$ gives
a rhombohedral polytype.
The example shown above has $\Delta=1$, and, as one may notice,
the cycle does not complete a period in $ABC$ notation: The cycle
is mapped to $(ACBCBCA)(B\ldots)(C\ldots)$ (starting from $A$).
This reflects the fact that the rhombohedral polytypes must be repeated
three times to complete the hexagonal unit cell \cite{verma_krishna}
(as indicated by the subscript $3$ in the Zhdanov symbol).
It is convenient to define two different periods $T_{01}$ and $T$:
$T_{01}$ is the period of a cycle in $01$ notation, and $T$ in $ABC$ notation.
Thus $T=T_{01}$ for $\Delta = 0$ and $T=3T_{01}$ for $\Delta =\pm1$.

As noted above, both $S$ and $I$ are good symmetries of the
Ising model as applied to polytypes.
We now want to address how these operators, defined in $01$ notation,
appear in the (somewhat more physical) $ABC$ notation.
We will use a different example, a SC $cyc=(101000001)$
(from $\hbox{}^{S+I} G_{7}$).

In 01 language the operator $S$ takes the form $0 \mathop{\iff}\limits^{S} 1$,
while space inversion $I$ is given by $(\sigma_1\sigma_2\ldots\sigma_N)
\mathop{\iff}\limits^{I} (\sigma_N\sigma_{N-1}\ldots\sigma_1)$.
Hence $S$ applied to $cyc$ is
\begin{equation}
\begin{array}{lllll lllll lllll llll}
&1\!\!\!&\!\!\!&0\!\!\!&\!\!\!&1\!\!\!&\!\!\!&0\!\!\!&\!\!\!&0\!\!\!&\!\!\!
&0\!\!\!&\!\!\!&0\!\!\!&\!\!\!&0\!\!\!&\!\!\!&1\!\!\!& \\
 A\!\!\!&\!\!\!&B\!\!\!&\!\!\!&A\!\!\!&\!\!\!&B\!\!\!&\!\!\!&A\!\!\!&\!\!\!&
 C\!\!\!&\!\!\!&B\!\!\!&\!\!\!&A\!\!\!&\!\!\!&C\!\!\!&\!\!\!&A
\end{array}
\;\stackrel{S}{\Longrightarrow}\;
\begin{array}{lllll lllll lllll llll}
&0\!\!\!&\!\!\!&1\!\!\!&\!\!\!&0\!\!\!&\!\!\!&1\!\!\!&\!\!\!&1\!\!\!&\!\!\!
&1\!\!\!&\!\!\!&1\!\!\!&\!\!\!&1\!\!\!&\!\!\!&0\!\!\!& \\
 A\!\!\!&\!\!\!&C\!\!\!&\!\!\!&A\!\!\!&\!\!\!&C\!\!\!&\!\!\!&A\!\!\!&\!\!\!&
 B\!\!\!&\!\!\!&C\!\!\!&\!\!\!&A\!\!\!&\!\!\!&B\!\!\!&\!\!\!&A
\end{array}
\ \ .
\end{equation}

As we can see here, the $S$ operation leaves one layer type invariant
(here arbitrarily chosen to be $A$), and takes $B \iff C$.

$I(cyc)$ is then
\begin{equation}
\begin{array}{lllll lllll lllll llll}
&1\!\!\!&\!\!\!&0\!\!\!&\!\!\!&1\!\!\!&\!\!\!&0\!\!\!&\!\!\!&0\!\!\!&\!\!\!
&0\!\!\!&\!\!\!&0\!\!\!&\!\!\!&0\!\!\!&\!\!\!&1\!\!\!& \\
 A\!\!\!&\!\!\!&B\!\!\!&\!\!\!&A\!\!\!&\!\!\!&B\!\!\!&\!\!\!&A\!\!\!&\!\!\!&
 C\!\!\!&\!\!\!&B\!\!\!&\!\!\!&A\!\!\!&\!\!\!&C\!\!\!&\!\!\!&A
\end{array}
\;\stackrel{I}{\Longrightarrow}\;
\begin{array}{lllll lllll lllll llll}
&1\!\!\!&\!\!\!&0\!\!\!&\!\!\!&0\!\!\!&\!\!\!&0\!\!\!&\!\!\!&0\!\!\!&\!\!\!
&0\!\!\!&\!\!\!&1\!\!\!&\!\!\!&0\!\!\!&\!\!\!&1\!\!\!& \\
 A\!\!\!&\!\!\!&B\!\!\!&\!\!\!&A\!\!\!&\!\!\!&C\!\!\!&\!\!\!&B\!\!\!&\!\!\!&
 A\!\!\!&\!\!\!&C\!\!\!&\!\!\!&A\!\!\!&\!\!\!&C\!\!\!&\!\!\!&A
\end{array}
\ \ .
\end{equation}
Thus, in $ABC$ notation, $I$ corresponds to the composite operation of
(spatial inversion)$\circ$($B\iff C$).

Given the above, it is clear that physically sensible Hamiltonians for
close-packed polytypes will be invariant under both $S$ and $I$
operations. We now proceed to examine the diffraction patterns of some
possible disordered ground states for Hamiltonians with $S+I$ symmetry.
Such `possible disordered ground states' are of course the D-pairs
obtained from $\hbox{}^{S+I} G_r$, as described above.

\section{Diffraction Patterns and Probabilities}

For the purpose of our study, we assume that the polytypic crystals
consist of unfaulted two-dimensional layers, stacked as prescribed by
the chosen SC of $\hbox{}^{S+I} G_r$.
Hence the diffracted intensity needs to be calculated only for
wavevectors normal to the close-packed layers (ie, along ${\bf c}$).
This problem has been addressed previously \cite{intensity,wilson,bw}
for perfect crystals and for various stacking defects; hence here we
only need apply old results to a novel kind of disorder.
The intensity of X-ray diffraction from close-packed crystals can
be expressed in terms of the number $N$ of layers in the chain, and
the average structure factor product $J(n)$, as \cite{bw}
\begin{equation}
   I(l)=N_{ab}\sum_{n=-N}^{N} (N-|n|)J(n)\exp(i2\pi nl).
   \label{int1}
\end{equation}
where $N_{ab}$ is a constant coming from a summation over the basal planes,
$l$ is a continuous variable which determines the wavevector
$k = 2\pi l /c$, and $n$ is an integer number of units of the primitive lattice
vector ${\bf c}$.
The average structure factor product $J(n)$ can be written as a
function of interlayer correlations, or probabilities, as follows.
Let $P_{AA}(n)$ be the probability that two layers $n$ apart be $A\cdots A$,
and similarly define $P_{AB}$ for $A\cdots B$, $P_{BA}$ for
$B\cdots A$, and so on.
With these probabilities $J(n)$ can be written as
\begin{eqnarray}
   J(n) &=&P_{AA}(n) F_A F_A^* +
           P_{BB}(n) F_B F_B^* +
           P_{CC}(n) F_C F_C^* +
           P_{AB}(n) F_A F_B^* +
           P_{BC}(n) F_B F_C^* + \nonumber \\
        & &P_{CA}(n) F_C F_A^* +
           P_{BA}(n) F_B F_A^* +
           P_{CB}(n) F_C F_B^* +
           P_{AC}(n) F_A F_C^*.  \label{jn}
\end{eqnarray}
The structure factors $F_A$, $F_B$ and $F_C$ for the hexagonal
$A$, $B$ and $C$ layers are also well-known\cite{intensity}.
They differ from each other only in phase since the layers represent
identical structures related by a rotation.
$F_B$ and $F_C$ can be written in term of the normalized
structure factor of $A$ layer ($F_A=1$) as
\begin{eqnarray}
     F_B &=  \exp [i2\pi m_0/3] \nonumber \\
     F_C &=  \exp [-i2\pi m_0/3]  \label{fabc}
\end{eqnarray}
where $m_0 = h_0 - k_0$ is an integer constant determined by the
components ($h_0,k_0$) parallel to the layers..
For $m_0 = 3m$, $m$ any integer, the structure factors are all unity, and,
as we can see from Eq. \ref{jn}, $J(n)$ does not depend on
the probabilities at all. Hence for this case, by Eq.\ \ref{int1},
the intensity is zero.
Taking $m_0=3m+1$ and inserting $F$ values (the case
$m_0=3m-1$ is trivially related), $J(n)$ reduces to
\begin{eqnarray}
   J(n) &=& P_{AA}(n) + P_{BB}(n) + P_{CC}(n) +
           [P_{AB}(n) + P_{BC}(n) + P_{CA}(n)]\exp(-i2\pi/3) + \nonumber \\
       & & [P_{BA}(n) + P_{CB}(n) + P_{AC}(n)]\exp( i2\pi/3).  \label{jn2}
\end{eqnarray}
Eq.(\ref{jn2}) shows that the intensity of the diffraction pattern depends
on the sums of the probabilities that two layers $n$ units apart are in
$A\cdots A$, in $A\cdots B$ or in $A\cdots C$ relationship.
After some algebra, using the fact $P_{AB}(n)=P_{BA}(-n)$, the intensity
$I(l)$ reduces to
\begin{equation}
   I(l) = {\sin^2(\pi Nl)\over sin^2(\pi l)} -
          2\sqrt{3}\sum_{n=1}^N (N-n)[Q_c(n)\cos(2\pi nl+{\pi\over 6})
          +Q_r(n)\cos(2\pi nl-{\pi\over 6})]   \label{int2}
\end{equation}
where $Q_c(n)\equiv P_{AB}(n) + P_{BC}(n) + P_{CA}(n)$ and
      $Q_r(n)\equiv P_{BA}(n) + P_{CB}(n) + P_{AC}(n)$.
[Here we use $c$ for `cyclic' and $r$ for `reverse'.
We also define $Q_s(n)=P_{AA}(n) + P_{BB}(n) + P_{CC}(n) = 1 - (Q_c(n) +
Q_r(n))$ ($s$=`same') for future use.]
Thus,
in order to calculate the diffracted intensity,
one needs only the lumped probabilities $Q_c(n)$ and $Q_r(n)$.

For perfectly periodic crystals these probabilities are
periodic, giving $\delta$-function peaks in the diffraction pattern.
On the other hand, for disordered crystals, the periodicity of
the probabilities may (or may not) be destroyed, depending on the type
of disorder which is introduced. If the correlations are not periodic,
then the diffraction pattern is of course continuous.

Previous work (that we are aware of) on disorder in close-packed
polytypes has concentrated on the random introduction, with various
probabilities, of various kinds of stacking faults in otherwise perfect
structures \cite{wilson,bw}. This approach gives nonperiodic probabilities
($Q_c$, $Q_r$) which decay to 1/3 at large $n$.

The disorder we are dealing with however is different: the only type of
`stacking fault' we consider is a zero-energy fault consisting of a free
choice among multiple paths in $G_r$, at one or more points
in an otherwise completely deterministic stacking sequence.
Furthermore, the distinct paths considered are always related by
symmetry. Our unfaulted or reference configuration, an
ordered or periodic polytype, is formed by choosing only one cycle, among
the two or more symmetry-related ones, to form an infinite chain.
Hence the earlier methods for computing the probabilities
are not appropriate for our case.
We found instead a simple rule for calculating the probabilities
$Q_x$ ($x=s,c,r$) for both the ordered and disordered close-packing sequences.
Some useful properties of these correlations are proved in the Appendix.

We now examine the correlations $Q_x$, comparing those for a perfect crystal
with those for a disordered chain built from a D-pair.
Fig.\ 2(a) shows $Q_s(n)$ for a perfect ($\Delta = -1$)
stacking sequence with $T_{01}=14$.
We can see that the period is $T=3T_{01}=42$.
$Q_s(n)$ and $Q_r(n)$ very similar, being related to $Q_s(n)$ by a shift in
$n$.

We now consider the disordered case.
Disordered chains were constructed by computer, using a
pseudo random number generator.
For type I and II D-pairs, this involves starting from a shared node $\cal N$
[for example see Fig.\ 1(a), 1(b)] and choosing one cycle
(either $cyc$ or $\mathop{cyc}\limits^{\longleftarrow}$ for type I, or $cyc$ or
$\mathop{\overline{cyc}}\limits^{\longleftarrow}$ for type
II) randomly; adding the chosen cycle brings us back to $\cal N$, and the
choice is made again.
For type III and IV D-pairs, four half cycles share
two nodes [Fig.\ 1(c), 1(d)].
There are two possible half cycles to be selected at each node;
again the choice is made randomly, with equal probability.
Given the symmetries of the problem, we believe that the
assumption of equal probabilities is reasonable for a real physical system.
In this manner a long disordered chain (over $50,000$ layers)
is produced, and its diffraction pattern computed from the correlations.

In the disordered case the value of $\Delta$ plays an
important role in determining the properties of the correlations.
Suppose the value of $\Delta$ of a cycle, say $cyc$, is $+1$ or $-1$,
and let the number of $1$'s ($0$'s) be $n_1$ ($n_0$).
In the cycle $\mathop{cyc}\limits^{\longleftarrow} = I(cyc)$, $n_1$
and $n_0$ will remain unchanged, so that $\Delta$ remains unchanged.
On the other hand, if the other half of the D-pair is
$\mathop{\overline{cyc}}\limits^{\longleftarrow} = SI(cyc)$, $n_1$ and
$n_0$ of $cyc$ are exchanged, so that
the value of $\Delta$ is switched to $-\Delta$.

When a disordered chain is built up from $cyc$ and
$\mathop{cyc}\limits^{\longleftarrow}$ [Fig.\ 1(a)],
the $\Delta$s of both cycles are the same.
Somewhat surprisingly, the result, as shown in
Fig.\ 2(b), is that the probabilities $Q_s(n)$ are
{\it periodic}, with period $T$ (recall that $T=3T_{01}$ for $\Delta = \pm1$
and $T=T_{01}$ for $\Delta = 0$), when $n$ is greater
than a threshold value $n_c$ (see Appendix).
It turns out that $n_c$ is a small number which is always less
than $T_{01}$, but the `noise' in this small
region makes a bump in the intensity (see below).
When $n>n_c$ we can expect (see Appendix for details)
that the basic properties of periodic correlations
can be applied for this case, so that the diffraction patterns
are very similar to those for the periodic case.
In particular, the locations of the $\delta$ functions are unchanged.

The correlations of a disordered chain grown by using
D-pairs of types II--V show a special behavior when $\Delta =\pm1$.
For example, a chain produced by a random mixture of
$cyc$ and $\mathop{\overline{cyc}}\limits^{\longleftarrow}$ (type II)
gives an arbitrary mixture of cycles with $\Delta =1$ and $-1$.
The other D-pairs (types III--V) also have this feature, as may be seen
from Fig.\ 1.
In the Appendix we show that the correlations, in these cases,
decay exponentially and approach $1/3$ [Fig.\ 2(c)]
(note that the same result was obtained
by Wilson\cite{wilson} using difference
equations for disordered hcp polytypes).
However when $\Delta$ is $0$ for any of types II--V,
the result is similar as that for a
$cyc$-$\mathop{cyc}\limits^{\longleftarrow}$ D-pair:
the chain has periodic probabilities for $n>n_c$ [with however
$Q_c(n)=Q_r(n)$], and irregular 'noise' below $n_c$.

\section{Diffraction Patterns}

After growing a close-packed but disordered
crystal from a D-pair,
we calculate the probabilities $Q_s(n)$, $Q_c(n)$ and
$Q_r(n)$,
and then calculate the intensity diffracted from these structures
by feeding these probabilities to Eq. (\ref{int2}).

Fig.\ 3 shows the intensity from a perfect crystal for various $\Delta$.
As noted above,
the perfect crystals are constructed by repetition of one cycle of
a given D-pair.
For $\Delta =0$ where $T=T_{01}$, in general, all $T$ diffraction lines
occur at $l = h/T$ with $h$ an integer.
Figure 3(a) shows the peaks in the range $0\le l\le 1$.
We see in contrast from Fig.\ 3(b) and 3(c) that $\delta$ function
intensities occur only at the positions $h=Tl = 3k+1$ for $\Delta =1$ and
$3k-1$ for $\Delta=-1$.
Then the number of peaks in the range $1\le h\le T$ is $T_{01} =T/3$.
These are well-known results; we only reproduce them here to facilitate
the comparison with the disordered cases.
The extinction of 2/3 of the peaks, for $\Delta=\pm 1$, is
proved in the Appendix.

In Fig.\ 4 we show the intensity diffracted from disordered
lattices.
When the disordered lattice is type I with $\Delta=0$ [Fig.\ 4(a)],
$\Delta=1$ [Fig.\ 4(b)] and $\Delta=-1$ [Fig.\ 4(c)],
we can see that the sharp peaks occur at the same positions
[compare, for example, Figs.\ 3(a) and 4(a)] as for the periodic cases,
but the intensities of the lines change.
We note also that there occurs a small intensity bump on the base
line of the patterns. This is the only evidence of the disorder in the
chain; we see that it can in fact be quite small, and hence difficult to
detect experimentally.

There are two reasons why the disorder is so well hidden in these cases.
One is of course the identity of $\Delta$ for the two halves of the
D-pair, which preserves long-range correlations as noted above.
The second reason is that the two cycles (say $cyc$ and
$\mathop{cyc}\limits^{\longleftarrow}$) must share at least one node,
and so have at least $r$ bits which are identical.
This leaves at most only ($T_{01} - r$) bits which can differ between
the two cycles. The actual number of differing bits can be as small as
one; hence the amplitude of the continuous part of the spectrum can be
quite small.

When the disordered lattice is built from type II--V D-pairs the
diffraction patterns can show very different behavior.
For $\Delta = 0$, the result is similar to that of type I:
all $T$ sharp peaks occur at the regular positions, with
an intensity bump due to the irregular part of the correlations [Fig.\ 5(a)].
[One difference is that, for this case, the intensity lines are
symmetrically placed about the axis $l=0.5$,
due to the fact that $Q_c(n)=Q_r(n)$].

Fig.\ 5(b) shows that some structures, which are possible ground states
of a class of symmetric Hamiltonians,
show a completely diffuse diffracted intensity with no $\delta$-function peaks.
When the D-pair is of type II through V, with $\Delta=\pm 1$,
the sharp $\delta$-function lines
are destroyed by the random mixture of cycles (or parts
of cycles) related by $SI$ (types II--V), or simply by $S$ (types III--V).
This can perhaps be understood in an intuitive way as follows.
Suppose two cycles, constituting the D-pair, have different $\Delta$
values: one has $+1$ and the other has $-1$.
Then, of course, if we have a perfect crystal (ie, all one cycle)
the sharp peaks
tend to occur at (respectively) $h=3k+1$ and $h=3k-1$.
Mixing the two cycles arbitrarily, the sharp
peaks of one cycle (say $cyc$) will be destroyed by any significant
fraction of the other cycle
(say, $\mathop{\overline{cyc}}\limits^{\longleftarrow}$)
since the positions of lines of $cyc$ corresponds
to the positions of zero intensities of
$\mathop{\overline{cyc}}\limits^{\longleftarrow}$.

Fig.\ 5(b) is for the case of an equal mixture of the two cycles.
In Fig.\ 6 we vary the ratio of the two cycles, in order to clarify how
the disordered, continuous pattern is related to the two discrete
spectra obtained for the pure periodic cases
($cyc$ and $\mathop{\overline{cyc}}\limits^{\longleftarrow}$).
We see from Fig.\ 6 that, as claimed above, the intensity
is changed from discrete to continuous by any finite fraction
of the symmetry-related cycle. It is also clear that the peaks
are smoothly shifted, as a function of the mixture ratio,
from one limit ($3k+1$) to the other ($3k-1$).
(We are of course not aware of any physical mechanism which would
bias the ratio away from 1/2; we include Fig.\ 6 simply to clarify the
behavior of the continuous spectra for these types of D-pairs.)

\section{D-pairs}

We have compiled a modest catalog of the D-pairs which may occur
for $r=6$ and $r=7$, always assuming $S+I$ symmetry. To do this,
we drew the graphs $\hbox{}^{S+I} G_{6}$ and $\hbox{}^{S+I} G_{7}$
by hand, and picked out by inspection the simple cycles
which satisfy the CW rules for D-pairs \cite{cw}.
We also found a few D-pairs for $r=8$, since this is the smallest $r$
giving type V D-pairs for $S+I$ symmetry.
Our method rapidly becomes cumbersome for larger $r$, and is already
rather unwieldy for $r\ge 7$. Hence, if any further search for
D-pairs at larger $r$ is warranted, it should be automated
(which is possible) and carried out by a machine.

Our manual search is however extensive enough to uncover
both types of D-pair in $\hbox{}^{S+I} G_r$ found by CW.
Therefore, we believe that the five types of Fig.\ 1 represent
all the topological types of D-pairs that can occur in $G_r$.
Hence we feel that the present study has revealed all the qualitative
features of disorder which may occur for Ising Hamiltonians.

In Table I, we list all the D-pairs we found from $\hbox{}^{S+I} G_{6}$.
Table I is actually not typical of larger $r$, since $r=6$ is the smallest
value, given $S+I$ symmetry, for which D-pairs occur at all.
However the D-pairs for larger $r$ are very numerous; hence we just
summarize those results here.

For $r=7$ we found 66 D-pairs, of which 38 gave $\delta$-function patterns and
28 gave pure continuous spectra. The periods $T_{01}$ for the $\delta$-function
spectra included 9--15, 17--19, 22, 24, 26, 28, and 30; for the
continuous spectra we found periods 9, 10, 14, 16, 18, 20, 22, 26, 28,
30, 34, and 38. There are no type V D-pairs for $r=7$,
since $\hbox{}^{S+I} G_{7}$ $\sim \hbox{}^IG_6$ \cite{cw}.
We did find (with help from a computer search)
type V D-pairs in $\hbox{}^{S+I} G_{8}$; typically,
they involve nonzero $\Delta$ values and so give continuous patterns.
Hence the principal feature which distinguishes
this type (besides their topological structure in the graph) is their
higher entropy per $T_{01}$ layers (which may or may not yield a higher
entropy per layer, depending on $T_{01}$).

We note finally that, for $r=6$ or 7,
odd-period D-pairs with continuous diffraction patterns
are rare. The reason is that such patterns are only obtained, for these $r$
values, from
type II pairs with nonzero $\Delta$. Type I pairs always give $\delta$-function
patterns, and types III--V always have even period (because all these types,
for $r<8$, are $S$-symmetric, ie, they `unfold'
into two doubled cycles rather than four distinct ones). From Table I,
the continuous spectra with odd period do not appear to be uncommon; however,
they represent only 2 out of 66 D-pairs for $r=7$.

\section{Discussion and Summary}

Recent work by Canright and Watson (CW) \cite{cw} has proposed
a novel type of disordered ground state for classical one-dimensional
chains whose Hamiltonians obey certain symmetries.
This disorder involves arbitrary mixtures of simple sequences (cycles)
which are related by symmetry.
Our goal in the present work has then been to ask how this kind of
disorder might appear in an experimentally accessible signal.
We chose to view the 1D chains as stackings of identical layers, ie, as
polytypes, and computed the the diffraction patterns along the stacking
direction.
In particular, we assumed, as appropriate for close-packed polytypes,
that the distinct ways of stacking the layers gave simple phase shifts in the
scattering.

Given these assumptions, we found that the diffraction patterns fell into
two classes: discrete, with a continuous baseline, or pure continuous.
In the former case the baseline can be quite small
[see for example Fig.\ 5(a)], giving a pattern strongly similar
to that for a periodic structure; or it can be larger [Fig.\ 4(a)].
In the latter case [Fig.\ 5(b)],
the disorder is obvious; however the remnant periodicity is also clear.
Briefly (see the Appendix for details), the reasons for the two
classes are as follows. In binary notation, there is long-ranged order
(LRO) for all types of
D-pairs, because the shared nodes (which represent at least $r$ bits)
recur with perfect periodicity. However, when the binary chain is
translated to a close-packed $ABC$ sequence, this LRO may or may not be lost,
depending on the relative symmetries of the pieces which are mixed
by the degeneracy, and on the parameter $\Delta$ which determines the net
shift in spatial phase after one period.

These results, being based on an entire class of generic Hamiltonians,
are as yet purely theoretical. It remains to be seen whether or not
real materials
might be governed by any of that fraction of Hamiltonians which give disordered
ground states. Such Hamiltonians should apply to any compound whose
structure consists of stacked identical layers, with the layers
restricted to a discrete number (in this paper, two) of states
(say, orientations). Here we have assumed the simplest of such structures,
ie, close-packed polytypes.

There are many well-known polytypic materials whose structures obey the
$ABC$ state restriction. SiC, perhaps the most well-studied case,
does not look promising \cite{verma_krishna,medium}, for two reasons.
First, the effective interlayer potentials \cite{medium} are likely to
have a periodic ground state, because they fall off too rapidly
(being very small for
$r\vbox {\hbox{\lower 0.9\baselineskip \hbox{$>$}} \break
         \hbox{\lower 0.2\baselineskip \hbox{$\sim$}} } 5$).
It is of course of interest to
see whether these potentials are close to any of those which give disordered
ground states. The answer to this question however requires further work;
specifically, the `inverse problem' of finding the set of Hamiltonians
which correspond to a given ground state (eg, a D-pair) must be solved.
We plan to address this question in future work.

Besides the `classical' polytypes such as SiC, there are polytypic structures
among simple metals and metallic alloys \cite{long}. Here the effective
potentials are long-ranged, and highly frustrated. The approach of CW
\cite{rs,teubner,cw} is strictly valid only for finite-ranged interactions.
However there remains the possibility that the ground state for the long-ranged
interactions is already determined by those up to some range $\hat{r}$,
so that our logic may be useful even for this case. Zangwill and Bruinsma
\cite{zangwill} have argued that the ground states for this class of problem
form a `devil's staircase' as a function of the average valence $Z$
of the metallic
alloy. This argument implies a periodic ground state for all $Z$.
This result depends on allowing elastic displacements of the layers from
perfectly periodic $\bf c$-axis spacing; this feature is also absent from
the simple discrete state space considered here. Thus, it seems that
the best current logic predicts periodic ground states for metallic polytypes;
however, there is clearly room for, and need for, further work
on this question.

Our theoretical approach to polytypes, and many other approaches, share the
common feature of reducing a three-dimensional, quantum-mechanical problem to
a one-dimensional classical problem. We believe such an approach is well
justified by the large mass of the layers. However, one must consider
the possibility that true quantum-mechanical solids \cite{qm} may not sample
the entire range of possibilities of classical interlayer Hamiltonians, even
after allowing for symmetry. This question also we plan to address in
future work.

In summary, our study of the diffraction patterns of D-pairs is motivated
by an interest in determining whether the kind of disordered ground states
identified by Canright and Watson \cite{cw} may arise in real materials.
We find that, depending on the details of the disordered structure,
a D-pair may give either a diffraction spectrum ($\delta$ functions)
very much like that for a perfect crystal (with however some continuous
spectrum as well); or one may find a purely continuous spectrum, in
which the underlying disorder is obvious. These results are an essential
component of any search for disordered ground states in polytypes.

\noindent{\bf Acknowledgements.} We thank Greg Watson for
help in finding $r=8$ D-pairs by a computer search.
This work was funded by the NSF under Grant \# DMR-9413057.
\pagebreak

\appendix
\section{}

\subsection{Probabilities for the Periodic Chain}

In this Appendix we prove various properties of the probabilities $Q_x$,
which are useful for understanding the diffraction patterns. We begin,
in this section, with some general properties for the periodic case.
The following sections
will then consider the presence or absence of sharp lines for
the periodic and disordered (D-pair) cases, respectively.

Define $\Delta (l)$ to be the $\Delta$ of a set of $l$ spins within
an interval, from the
$(q+1)$st to the $(q+l)$th spin ($q,l$ integers with $q\ge 0$
and $l\ge 1$) in an infinite Ising chain.
(Hence $\Delta(l) = \Delta_q(l)$ strictly depends on $q$ as well; but we
will usually ignore the $q$ dependence.)
The $\Delta $ defined in the main text is then $\Delta(T_{01})$ in this
notation.
Also define $n_s(l)$, $n_c(l)$ and $n_r(l)$ to be the
number of sets which give $\Delta(l)=0,+1$, and $-1$ respectively when we scan
$q$ from $0$ to $T_{01}-1$.
Then, for the periodic case,
\begin{eqnarray}
     Q_s(mT+n) &=& {n_s(n)\over T_{01}} \nonumber \\
     Q_c(mT+n) &=& {n_c(n)\over T_{01}} \label{prob_periodic}\\
     Q_r(mT+n) &=& {n_r(n)\over T_{01}} \nonumber
                    \quad\quad (m\ge 0) \nonumber
\end{eqnarray}

We can prove these relations as follows.
First consider a set of $n$ spins and a corresponding close-packing sequence.
If the stacking sequence begins with $A$ $(B,C)$ then after $n$
further layers the position will be
$A$ $(B,C)$ if $\Delta(n)=0$, or will be $B$ $(C,A)$ if $\Delta(n) = +1$,
or $C$ $(A,B)$ if $\Delta(n)=-1$.
Thus, for example,
$n_s(n)$ is the number of pairs of layers, $n$ apart, which are in
the same (hence `$s$') orientation: $A\cdots A$, $B\cdots B$ or $C\cdots C$.
(Similarly, as in the main text,
`$c$' stands for `cyclic' and `$r$' for `reverse'.)

Let $N_p(n)$ be the number of layer-pairs, $n$ apart, over a whole chain, and
$n_s^{(\infty)}(n)$ be the number of pairs among $N_p(n)$ which gives
$\Delta(n) = 0$.
For a periodic chain composed of $N_u$ unit cycles with period $T_{01}$
we will have $N_p = T_{01}N_u$ and $n_s^{(\infty)}(n) = n_s(n)N_u$.
With this notation $Q_s(n)$ can be written as
\begin{equation}
     Q_s(n) = {n_s(n)\over T_{01}}
     \quad\quad (1\le n< \infty).\label{qmiddle}
\end{equation}
Since the chain is periodic it is useful to write $n$ as $n=mT+n'$,
$m\ge 0$ and $1\le n'\le T$.
Then $n_s(n)=n_s(mT+n') =n_s(n')$
[since $\Delta(mT+n')=\Delta(mT)+\Delta(n')=\Delta(n')$].
Finally we can drop the prime, giving
\begin{equation}
     Q_s(mT+n) = {n_s(n)\over T_{01}}
       \label{qfinal}
\end{equation}
Other equations in Eqs.\ (\ref{prob_periodic}) can be derived using
the same method.
Thus we do not need to count the number of layer-pairs
over a whole chain in order to calculate the probabilities
$Q_s(n)$, $Q_c(n)$ and $Q_r(n)$. Instead only one to three unit cycles
are enough in calculating the probabilities (see below).

In the remainder of this section
we will take $\Delta=+1$ for
concreteness. (Generalizations to other $\Delta$ are obvious.)
We want to derive relationships between $n_x(n)$ and $n_x(n')$, where
$n$ and $n'$ differ by a multiple of $T_{01}$. For example, let us
show that $n_r(T_{01}+n) = n_c(n)$ and
$n_r(2T_{01}+n) = n_s(n)$ for the case $\Delta =+1$.
We will use the modular arithmetic properties of
$\Delta(n)$:
\begin{equation}
      (\pm 1) + (\pm 1) = \mp 1 \quad\mbox{and}\quad
      (\pm 1) - (\pm 1) = 0. \label{delta}
\end{equation}
A set of $T_{01}+n$ spins contributes to $n_r$ ($n_c$, $n_s$)
if $\Delta(T_{01}+n)$ is $-1$ ($+1$, $0$).
Thus (for example)
$\Delta(T_{01}+n) = -1 = \Delta(T_{01})+\Delta(n)$ implies that
$\Delta(n)$ should be $+1$, since
$\Delta\equiv\Delta(T_{01})=+1$.
With this argument we can see that $n_r(T_{01}+n)=n_c(n)$.
Similarly, if $\Delta(2T_{01}+n$) is $-1$,
then $\Delta(n)=0$, since $\Delta(2T_{01})$ is $-1$ for the
case of $\Delta = +1$. Hence $n_r(2T_{01}+n)= n_s(n)$.
Finally we see that
\begin{eqnarray}
       Q_r(T_{01}+n) &=& Q_c(n) \nonumber \\
       Q_r(2T_{01}+n) &=& Q_s(n).
\end{eqnarray}

We can consider properties under subtraction, as well as under addition.
Noting that
$\Delta_q(T-n) = \Delta(T) - \Delta_{q-n}(n)$, and averaging to
eliminate the $q$ dependence, we find, for example,
\begin{equation}
     Q_r(n) = Q_c(T-n). \quad (1\le n\le T)
\end{equation}
This result is independent of $\Delta$ since it only uses the fact that
$\Delta(T) = 0$.

With the logic described above it is not hard to derive a number of relations
among the probabilities. In the next section we will use such relations
as needed.

\subsection{Extinction of Diffraction Lines}
Using the properties shown in section I, we can show
that the lines indexed $3k$ and $3k-1$ ($k\ge 0$) are extinguished
for a periodic chain with $\Delta = +1$.
Rewrite the intensity equation [Eq. (\ref{int2}] as
\begin{eqnarray}
   I(l)&=&{\sin^2(\pi Nl)\over \sin^2(\pi l)} -
          2\sqrt{3}\sum_{n=1}^N (N-n)[Q_c(n)\cos(2\pi nl+{\pi\over 6})
          +Q_r(n)\cos(2\pi nl-{\pi\over 6})]   \nonumber\\
       &\equiv&{\sin^2(\pi Nl)\over \sin^2(\pi l)} - {\cal S}. \label{aint2}
\end{eqnarray}
Using the fact that the probabilities $Q_c(n)$ and $Q_r(n)$
are periodic and that $Q_c(T-n)=Q_r(n)$ we can simplify the
summation in Eq. (\ref{aint2}) as
\begin{equation}
   {\cal{S}} \equiv {2\sqrt{3}N^2\over T}
             \sum_{n=1}^TQ_r(n)\cos(2\pi nl-{\pi\over 6}). \label{aint3}
\end{equation}
Next, using $Q_r(T_{01}+n)=Q_c(n)$ and
$Q_r(2T_{01}+n)=Q_s(n)$, ${\cal{S}}$ can be written as
\begin{eqnarray}
   {\cal{S}}&=&{2\sqrt{3}N^2\over T}
             \sum_{n=1}^{T_{01}}
            [Q_r(n)\cos (2\pi nl-{\pi\over 6})
            +Q_c(n)\cos (2\pi T_{01}l+2\pi nl-{\pi\over 6}) \nonumber\\
           &+&Q_s(n)\cos (4\pi T_{01}l+2\pi nl-{\pi\over 6})]. \label{aint4}
\end{eqnarray}
Let $l=h/T$ ($h$=integer).
The first term in Eq.\ (\ref{aint2}) is $0$ everywhere except
$h=0$ and $h=T$. Thus if ${\cal{S}}$ vanishes for any other $h$,
$0<h<T$, then the
intensity lines are extinguished.

(i) $h=0$ or $T$

${\cal{S}}=N^2$ and then the intensity $I = N^2-N^2=0$.

(ii) $h=3k$
\begin{equation}
   {\cal{S}} = {2\sqrt{3}N^2\over T}
             \sum_{n=1}^{T_{01}}[Q_r(n)+Q_c(n)+Q_r(n)]
             \cos({2\pi kn\over T_{01}}-{\pi\over 6}).
\end{equation}
The sum of all three probabilities is always $1$, and the
summation of the cosine term can be carried out to be $0$.
Thus $I(3k) = 0$.

(iii) $h=3k+2 = 3k-1$

Using $Q_c(n)=Q_s(T_{01}-n)$, the last two terms in Eq. (\ref{aint4})
cancel. Hence $\cal S$ becomes
\begin{equation}
   {\cal{S}} = {2\sqrt{3}N^2\over T}
             \sum_{n=1}^{T_{01}} Q_r(n)
             \cos \left ({2\pi n(3k-1)\over 3T_{01}}-{\pi\over 6}\right ).
\end{equation}
By straightforward algebra we can show the above equation
also vanishes, using $Q_r(n)=Q_r(T_{01}-n)$.
Hence $I(3k-1)=0$.

Thus we find that only one-third of the allowed lines appear in the
rhombohedral case $\Delta= +1$; the same conclusion holds for $\Delta =
-1$. In the hexagonal case ($\Delta=0$) all $T$ lines appear.

\subsection{Disordered Case}

For the disordered case, we need to consider the average
values for $n_s(n)$, $n_c(n)$ and $n_r(n)$ over
the entire chain. (For the remainder of this section we will use the
notation $n_x(n)$, earlier defined for periodic chains, to denote this
average.) For simplicity, we will assume that the chain is
built from a random mixture of two symmetry-related cycles, present with
equal probability. Hence our derivation will be strictly valid only for
type I and II D-pairs; however our conclusions (periodic probabilities
and $\delta$-function spectra for pairs with the same $\Delta$,
decaying probabilities and loss of sharp
diffraction lines for mixtures of cycles of opposing $\Delta$)
are valid in general.

We consider two cycles $c$ and $c'$, related by symmetry
($I$ or $SI$), with bit parities $\Delta$ and $\Delta'$, respectively.
We will need to distinguish the case that
$\Delta = \Delta'$ (`same-$\Delta$' case; type I D-pairs, or type II
with $\Delta =0$) from that with
$\Delta = -\Delta'$ (`opposing-$\Delta$' case; type II with $\Delta= \pm 1$).

Our strategy is to first calculate the short-ranged $Q$'s, then the
longer-ranged ones, building from the short-ranged values. For $n$
sufficiently large ($\ge 2T_{01}+1$), there is always at least one
complete cycle, of length $T_{01}$, which appears unchanged in all sums
involved in the averaging process used to compute the $n_x(n)$ and hence
the $Q_x(n)$ ($x=s,c,r$). We call such a cycle `lumped'.
If a cycle, or fraction of a cycle, is included but not lumped, we say it
is `scanned'.

(i) $n=1$

We begin with $n=1$.
For a same-$\Delta$ pair, $n_1$ and $n_0$ are the same for either $c$ or
$c'$. Thus it is clear that $Q_c(1)=n_1N_u/N=n_1/T_{01}$,
$Q_r(1)=n_0N_u/N=n_0/T_{01}$ and $Q_s(1)=0$.
These are the same as for the periodic case.
In the opposing-$\Delta$ case
the probabilities will depend on the probabilities of the two cycles.
Taking an equal fraction of the two cycles, the probabilities
(neglecting edge effects) are
$Q_c(1)=Q_r(1)=(1/2)(n_1+n_0)/T_{01}= 1/2$.

(ii) $2\le n\le T_{01}+1$

If $2\le n\le T_{01}+1$ the length of spin sets involved
in the counting process will be up to $2T_{01}$---that is, two cycles
are scanned
[from the first spin ($q=0$)
to $2T_{01}$th spin ($q=T_{01}-1$)]\cite{qrange}.
Since we scan more than one cycle in this case, we need to compute the
$n_x(n)$ by averaging over the four possible combinations of two cycles:
$cc$, $cc'$, $c'c$, $c'c'$.
We find that, in the same-$\Delta$ case,
there exists a threshold ($n_c$) between $1$ and $T_{01}$,
above which the probabilities $Q_x$ are periodic.
Since the most general bound is $n_c < T_{01}+1$, we will confine
ourselves to showing the periodicity for $n>T_{01}+1$.

(iii) $T_{01}+1 < n\le 2T_{01}$

For $n$ in this range, three unit cycles are scanned.
Therefore the $n_x(n)$
should be calculated by averaging over eight possible combinations
of the two cycles ($ccc$, $ccc'$, etc).
Let these averaged values be $\overline{n}_s^{(0)}(n)$,
$\overline{n}_c^{(0)}(n)$ and $\overline{n}_r^{(0)}(n)$.
Then the probabilities can be written as
\begin{eqnarray}
     Q_s(n) &=& {\overline{n}_s^{(0)}(n) \over T_{01}}\nonumber \\
     Q_c(n) &=& {\overline{n}_c^{(0)}(n) \over T_{01}}\label{prob_disorder}\\
     Q_r(n) &=& {\overline{n}_r^{(0)}(n) \over T_{01}} \nonumber
                    \quad\quad (T_{01}+1 < n\le 2T_{01}).
\end{eqnarray}

(iv) $2T_{01}+1\le n\le 3T_{01}$

In this region there exists one unit cycle which is lumped (ie, makes an
invariant addition to the sum)
when we scan $q$ from $0$ to $T_{01}-1$.
This cycle can be either $c$ or $c'$ with equal probability, and
the $\Delta(T_{01})\equiv \Delta$ of this cycle will be
$0$, $+1$ or $-1$ depending on the D-pair considered.
Let $f_1^{(s)}$, $f_1^{(c)}$ and $f_1^{(r)}$ be the probabilities
[we will use the term 'cycle-probability' in order to distinguish these
probabilities from the $Q_x(n)$] that the $\Delta$ of the lumped
cycle is $0$, $+1$, or $-1$ respectively.
For example, for a type II D-pair, in which the $\Delta$ of $cyc$ is $+1$
and that of $\overline{cyc}$ is $-1$, $f_1^{(s)}=f_1^{(r)}=1/2$ and
$f_1^{(s)}=0$.
Recalling the modular arithmetic of the $\Delta$'s,
the $n_x(n)$ in this range can be written in terms of the
cycle-probabilities and $\overline{n}_s^{(0)}(n)$, $\overline{n}_c^{(0)}(n)$
and $\overline{n}_r^{(0)}(n)$:
\begin{eqnarray}
 \overline{n}_s(2T_{01}+n) &=&
    f_1^{(s)}\overline{n}_s^{(0)}(n)+f_1^{(r)}
        \overline{n}_c^{(0)}(n)+f_1^{(c)}\overline{n}_r^{(0)}(n)\nonumber\\
 \overline{n}_c(2T_{01}+n) &=&
    f_1^{(c)}\overline{n}_s^{(0)}(n)+f_1^{(s)}
        \overline{n}_c^{(0)}(n)+f_1^{(r)}\overline{n}_r^{(0)}(n)
              \label{nbar1}\\
 \overline{n}_r(2T_{01}+n) &=&
    f_1^{(r)}\overline{n}_s^{(0)}(n)+f_1^{(c)}
        \overline{n}_c^{(0)}(n)+f_1^{(s)}\overline{n}_r^{(0)}(n).\nonumber
\end{eqnarray}
Dividing Eq. (\ref{nbar1}) by $T_{01}$,
we can calculate the probabilities $Q_s(n)$, $Q_c(n)$ and
$Q_r(n)$.

(v) Larger $n$

If $n$ is in the range $3T_{01}+1\le n\le 4T_{01}$,
there are two lumped cycles.
The cycle-probabilities $f_2^{(s)}$, $f_2^{(c)}$, and $f_2^{(r)}$
can then be obtained using those obtained before:
$f_2^{(s)}=f_1^{(s)}f_1^{(s)}+f_1^{(r)}f_1^{(c)}+f_1^{(c)}f_1^{(r)}$
and so on. From these cycle-probabilities we can calculate the $n_x(n)$
as before.

In general, if $n$ is in the range
$(p+1)T_{01}+1\le n\le (p+2)T_{01}$ ($p\ge 1$),
we can use the following formulae:
\begin{eqnarray}
 f_p^{(s)}&=&
    f_{p-1}^{(s)}f_1^{(s)}+f_{p-1}^{(r)}f_1^{(c)}+f_{p-1}^{(c)}f_1^{(r)}
    \nonumber \\
 f_p^{(c)}&=&
    f_{p-1}^{(c)}f_1^{(s)}+f_{p-1}^{(s)}f_1^{(c)}+f_{p-1}^{(r)}f_1^{(r)}
    \label{cycle_prob} \\
 f_p^{(r)}&=&
    f_{p-1}^{(r)}f_1^{(s)}+f_{p-1}^{(c)}f_1^{(c)}+f_{p-1}^{(s)}f_1^{(r)}.
    \nonumber
\end{eqnarray}
Since $f_1^{(s)}$, $f_1^{(c)}$ and $f_1^{(r)}$ can be determined by
the $\Delta$ of the D-pair, we can obtain the cycle-probabilities for any $p$
using these recursion relations.
Using Eqs.\ (\ref{cycle_prob}) we can finally get $n_x[(p+1)T_{01}+n]$:
\begin{eqnarray}
 \overline{n}_s((p+1)T_{01}+n) &=&
    f_p^{(s)}\overline{n}_s^{(0)}(n)+f_p^{(r)}
        \overline{n}_c^{(0)}(n)+f_p^{(c)}\overline{n}_r^{(0)}(n)\nonumber \\
 \overline{n}_c((p+1)T_{01}+n) &=&
    f_p^{(c)}\overline{n}_s^{(0)}(n)+f_p^{(s)}
        \overline{n}_c^{(0)}(n)+f_p^{(r)}\overline{n}_r^{(0)}(n)
    \label{nbar2}\\
 \overline{n}_r((p+1)T_{01}+n) &=&
    f_p^{(r)}\overline{n}_s^{(0)}(n)+f_p^{(c)}
        \overline{n}_c^{(0)}(n)+f_p^{(s)}\overline{n}_r^{(0)}(n)\nonumber\\
    & & (p\ge 2,\quad 1\le n\le T_{01}). \nonumber
\end{eqnarray}
Thus, once we know $\overline{n}_s^{(0)}(n)$,
$\overline{n}_c^{(0)}(n)$ and $\overline{n}_r^{(0)}(n)$,
and the $f_p^{(x)}$ from the recursion relations,
we can calculate the probabilities $Q_x((p+1)T_{01}+n)$ by dividing
Eqs.\ (\ref{nbar2}) by $T_{01}$.

When $\Delta = +1$ in a type I D-pair, for example, we know that
$f_1^{(s)}=0$, $f_1^{(c)}=1$, and $f_1^{(r)}=0$. The recursion relations
(\ref{cycle_prob}) for this case will be
\begin{equation}
        f_p^{(s)}=f_{p-1}^{(r)}\quad f_p^{(c)}=f_{p-1}^{(s)}
  \quad f_p^{(r)}=f_{p-1}^{(c)}\ \ ,
\end{equation}
so that the probabilities are given by

\begin{eqnarray}
  Q_s(2T_{01}+n)={\overline{n}_r^{(0)}(n)\over T_{01}}&\quad\quad
 &Q_s(3T_{01}+n)={\overline{n}_c^{(0)}(n)\over T_{01}} \quad\quad
  Q_s(4T_{01}+n)={\overline{n}_s^{(0)}(n)\over T_{01}} \nonumber \\
  Q_r(2T_{01}+n)={\overline{n}_c^{(0)}(n)\over T_{01}}&
 &Q_r(3T_{01}+n)={\overline{n}_s^{(0)}(n)\over T_{01}} \quad\quad
  Q_r(4T_{01}+n)={\overline{n}_r^{(0)}(n)\over T_{01}} \\
  Q_c(2T_{01}+n)={\overline{n}_s^{(0)}(n)\over T_{01}}&
 &Q_c(3T_{01}+n)={\overline{n}_r^{(0)}(n)\over T_{01}} \quad\quad
  Q_c(4T_{01}+n)={\overline{n}_c^{(0)}(n)\over T_{01}}. \nonumber
\end{eqnarray}
Thus, even for the disordered chain,
we can see that the probabilities are periodic, when $n>n_c$,
with  period $3T_{01}\equiv T$ for the $\Delta = +1$ case.
In that region of $n$ the properties of periodic probabilities
(see above) can be applied to this disordered case.
Therefore we can expect that the diffraction lines ($\delta$ function)
in Fourier space occur at the same positions as those for a periodic chain,
as derived above.
With the same method we can show the periodicity of probabilities
for any same-$\Delta$ case, ie, any type I D-pair, or for a type II
D-pair with $\Delta =0$. It is clear that the proof generalizes to types
III--V, as long as $\Delta=0$ for all possible cycles which can be
formed.

On the other hand, for the opposing-$\Delta$ cases
(type II--V D-pairs, with $\Delta\ne 0$)
we have a different story.
The initial cycle-probabilities for this case are different from the
previous example: they are $f^{(s)}_1=0$ and $f^{(c)}_1=f^{(r)}_1=1/2$. From
Eqs. (\ref{cycle_prob}) and the initial cycle-probabilities
it is not hard to get the following recursion relation,
\begin{equation}
   f_p^{(x)}={1\over 2}(f_{p-1}^{(x)} + f_{p-2}^{(x)}) \label{recur}
\end{equation}
(where again $x = s,c,r$);
here we have used the fact that $f_p^{(c)}=f_p^{(r)}$.
We can solve (\ref{recur}) to get
\begin{equation}
  \delta_p \equiv f_p - f_{p-1} = C\Bigl(-{1\over 2}\Bigr)^p
\label{deltap}
\end{equation}
with $C= 4(f_2-f_1)$. We then sum (\ref{deltap}) to get,
for large $p$,
\begin{equation}
  f_{p\to\infty}^{(x)}=f_1^{(x)}+{2\over 3}(f_2^{(x)}-f_1^{(x)})\label{largep}.
\end{equation}
With initial conditions appropriate to $\Delta = \pm 1$ (ie,
$f_1^{(s)}=0$ and $f_2^{(s)}=1/2$, or equivalently,
$f_1^{(c,r)}=1/2$ and $f_2^{(c,r)}=1/4$)
we can show that all three cycle-probabilities
decay to $1/3$ at large $p$.
With Eqs.\ (\ref{nbar1}),
$\overline{n}_s(\mbox{large}\ n)$ can be found to be
$(1/3)T_{01}$, using the fact that $\overline{n}_s^{(0)}(n)+
\overline{n}_c^{(0)}(n)+\overline{n}_s^{(0)}(n)=T_{01}$.
Finally, the probabilities for large $n$ are
\begin{equation}
  Q_s(n)={\overline{n}_s(n)\over T_{01}}={1\over3}=Q_c(n)=Q_r(n).
\end{equation}
The diffraction pattern for the opposing-$\Delta$ case is thus diffuse,
without any $\delta$ functions.

\pagebreak
\begin{table}
\caption{D-pairs found in $\hbox{}^{S+I} G_{6}$.
The first column gives the D-pairs in binary notation.
The type II pairs are always $cyc$/$\mathop{\overline{cyc}}
\limits^{\longleftarrow}$.
The type III pairs [see Fig.\ 1(c)] can be decomposed in several ways;
here we give them as $cyc$/$\mathop{cyc}\limits^{\longleftarrow}$.
The (binary) periods of the paired cycles are given,
as these are apparent even in the disordered
diffraction patterns [see, eg, Fig.\ 5(b)]. The $\Delta$ value and type
determines whether the diffraction pattern has $\delta$ functions, or is
continuous.}
\begin{tabular}{cccccc}
r &D-pair          &$T_{01}$
  &$\Delta$ Values\tablenotemark[1]
  &Type     &Spectrum \\
\tableline
 6 &
 ${\begin{array}[c]{c}
 1 0 1 1 1 1 0 0 1 0 0 0 0 1 \\
 0 0 1 1 1 1 0 1 1 0 0 0 0 1
\end{array}}$
& 14 &$0,0,-1,+1$& III & continuous\\
 6 &
 ${\begin{array}[l]{l}
 1 0 1 1 1 1 1 0 0 1 0 0 0 0 0 1    \\
 0 0 1 1 1 1 1 0 1 1 0 0 0 0 0 1
 \end{array}}$
& 16 &$0,0,-1,+1$ & III &continuous\\
 6 &
 ${\begin{array}[c]{c}
 1 0 0 1 0 1 1    \\
 0 0 0 1 0 1 1
 \end{array}}$
&  7 &$+1,-1$ & II & continuous \\
 6 &
 ${\begin{array}[c]{c}
 0 0 0 1 0 1 0 0 1 1 1 0 1 0 1 1    \\
 1 0 0 1 0 1 0 0 0 1 1 0 1 0 1 1
 \end{array}}$
& 16 &$0,0,+1,-1$ & III &continuous\\
 6 &
 ${\begin{array}[c]{c}
 1 0 1 1 1 1 1 1 0 0 1 0 0 0 0 0 0 1    \\
 0 0 1 1 1 1 1 1 0 1 1 0 0 0 0 0 0 1
 \end{array}}$
& 18 &$0,0,-1,+1$ & III &continuous\\
 6 &
 ${\begin{array}[c]{c}
 1 0 0 1 0 1 0 1 1    \\
 0 0 0 1 0 1 0 1 1
 \end{array}}$
&  9 &$+1,-1$ & II &continuous\\
\end{tabular}
\tablenotetext[1]{Each entry has two or four values depending on the
                  type of D-pair. (see Fig. 1)}
\end{table}

\begin{figure}
\caption{Schematic drawings of the five topological types of D-pair,
assuming $S+I$ symmetry for the Ising Hamiltonian.
     (a) Type I : $cyc$ and $\mathop{cyc}
                  \limits^{\longleftarrow}\equiv I(cyc)$.
     (b) Type II : $cyc$ and $\mathop{\overline{cyc}}
                  \limits^{\longleftarrow}\equiv SI(cyc)$.
     (c) Type III: There are four possible cycles, sharing two nodes.
Here and in (d) (type IV), we have shown symmetric D-pairs which are
invariant
under $S$, because such symmetry holds for these types for $r<8$,
which represents the majority of cases studied in this work.
     (d) Type IV : There are four possible cycles sharing two nodes.
     (e) Type V : There are sixteen possible cycles sharing four nodes.
Type V D-pairs are invariant under any combination of $S$ and $I$.}
\end{figure}

\begin{figure}
\caption{The probabilities $Q_s(n)$ for $T=42$.
     (a) For a perfect chain with $\Delta =-1$.
Other probabilities are related to $Q_s(n)$ by $Q_r(n)= Q_s(14+n)$
and $Q_c(n) = Q_s(28+n)$.
     (b) For a disordered chain; type II D-pair with $\Delta=0$.
         We can see the irregular part only for the first few $n$.
         In general, the probabilities are periodic for this
         (`same-$\Delta$') type of D-pair, for $n>n_c$, with $n_c <T_{01}$.
     (c) For a disordered chain; type II D-pair with $\Delta=\pm 1$.
         The probability decays exponentially to $1/3$.
         Others ($Q_c$, $Q_r$) show the same behavior.}
\end{figure}

\begin{figure}
\caption{Diffraction patterns from perfect chains with various $\Delta$.
     (a) For $\Delta = 0$, $T=T_{01}=13$.
         Basically all lines occur; the number of lines is $T$.
         Some lines are too small to see.
     (b) For $\Delta = +1$, lines at $l=h/T$, with $h = 3k$ and $h=3k-1$,
         are extinguished.
     (c) For $\Delta = -1$, $3k$ and $3k+1$ lines are extinguished.}
\end{figure}

\begin{figure}
\caption{Diffraction patterns from disordered chains (type I) with
        various $\Delta$.
     (a) Pattern for a disordered chain built from a D-pair (shown);
         the pattern for a perfect chain from half of this D-pair is
         shown in Fig.\ 3(a).
         Relative to Fig.\ 3(a), the positions of the lines are not
         changed; but there is a smooth background, with a visible `bump'.
     (b) $\Delta = +1$; compare with Fig.\ 3(b).
     (c) $\Delta = -1$; compare with Fig.\ 3(c).}
\end{figure}

\begin{figure}
\caption{Diffraction patterns from
        disordered chains built from a type II D-pair.
        The lines are symmetric with repect to $l=0.5$, due to
        $Q_c(n) = Q_r(n)$.
        (a) $\Delta =0$. Here we still have a same-$\Delta$ pair, and hence
        sharp lines in the spectrum.
        (b) When $\Delta\ne 0$ we have an `opposing-$\Delta$' pair.
        [An example of probabilities for an opposing-$\Delta$ case is
        shown in Fig.\ 2(c).]
        In this case, as shown in the Appendix, the probabilities are
        not periodic, and there are no sharp lines in the diffraction
spectrum.}
\end{figure}

\begin{figure}
\caption{The diffracted intensity with varying probabilities (as shown)
        for mixing the two cycles,
        $cyc$ and $\mathop{\overline{cyc}}\limits^{\longleftarrow}$,
from a type II D-pair with $\Delta =\pm 1$.
        We can see the positions of maximum intensity moving
        from $3k-1$ to $3k+1$ as the fraction
of $\mathop{\overline{cyc}}\limits^{\longleftarrow}$ increases (top to bottom).
        We have drawn a few dotted lines as a guide to the eye.
        Fig.\ 5(b) shows the case of a 50:50\% mixture.}
\end{figure}


\begin{references}
\bibitem{rs} Charles Radin and Lawrence S.~Schulman, Phys. Rev. Lett.
{\bf 51}, 621 (1983).

\bibitem{teubner} Max Teubner, Physica A {\bf 169}, 407 (1990).

\bibitem{bruijn} N. G. de Bruijn, Koninklijke Nederlands Akademie van
                 Wetenschappen, Proc. {\bf 49}, 758 (1946).

\bibitem{cw} Geoff Canright and Greg Watson, unpublished.

\bibitem{3law} Jacek Mi\c ekisz and Charles Radin, {\sl Mod.\
Phys.\ Lett.\ B} {\bf 1}, 61 (1987), and references therein.

\bibitem{verma_krishna} See, for example, Ajit Ram Verma and P.~Krishna,
{\sl Polymorphism and Polytypism in Crystals}, John Wiley \& Sons
(1966).

\bibitem{entropy} Note that, while the entropy per layer is finite,
the entropy {\it per atom\/} is of ${\cal O}(N_{atom}^{-2/3})$.

\bibitem{crutch} J.P.\ Crutchfield and K.\ Young, Phys.\ Rev.\ Lett.\
{\bf63}, 105 (1989).

\bibitem{hagg} Gunnar H\"agg, Arkiv. F\"or Kemi,
   Mineralogi Och Geologi {\bf 16B} (1943) 1-6.

\bibitem{zhdanov} G. S. Zhdanov.
 Compt. Rend. Acad. Sci. URSS {\bf 48} (1945) 43.

\bibitem{intensity}
 M. S. Paterson, J. Appl. Phys. {\bf 23}, 805 (1952);
 T. R. Welberry, Rep. Prog. Phys. {\bf 48}, 1543 (1985);
 A. Guinier, {\it X-Ray Diffraction} (W. H. Freeman and Co.,
              San Francisco and London, 1963).

\bibitem{wilson} A. J. C. Wilson, Proc. R. Soc. London Ser. A
                 {\bf 180}, 277 (1941).

\bibitem{bw}R. Berliner and S. A. Werner, Phys. Rev. B
            {\bf 34}, 3586 (1986) and references therein.

\bibitem{medium} C. Cheng, R.J. Needs, V. Heine, and N. Churcher,
{\sl Europhys. Lett.} {\bf 3}, 475 (1987);
C. Cheng, R.J. Needs, and Volker Heine, {\sl J. Phys. C} {\bf 21},
1049 (1988);
V. Heine, in {\sl Competing Interactions and Microstructures:
Statics and Dynamics},R. LeSar, A. Bishop, and R. Heffner. eds.
(Springer-Verlag, Berlin, 1987).

\bibitem{long} A. Blandin, J. Friedel, and G. Saada,
{\sl J. Phys. (Paris) Colloq.} {\bf 27}, C3-128 (1966);
Charles W. Krause and J.W. Morris, Jr., {\sl Acta Met.} {\bf 22}, 767 (1974).

\bibitem{zangwill} A.\ Zangwill and R.\ Bruinsma, Comm.\ Cond.\
Matt.\ Phys.\ {\bf 13}, 1 (1987); R.\ Bruinsma and A.\ Zangwill,
Phys.\ Rev.\ Lett.\ {\bf 55}, 214 (1985).

\bibitem{qm} For a recent quantum-mechanical discussion of
one-dimensional disorder in layered materials, see Ping Sheng and
Z.-Q.\ Zhang, Phys.\ Rev.\ Lett. {\bf 74}, 1343 (1995). For a review
see R.\ Merlin, IEEE J.\ Quantum Elect.\ {\bf 24}, 1791 (1988).

\bibitem{qrange} For simplicity we take $q=0$ to be the first spin in a
                 unit cycle.
                While the details of the argument depend on this choice,
the conclusions do not.

\end{references}
\end{document}